\documentclass{elsart}
\usepackage{graphicx,epsfig}

\begin{document}
\def\ltsimeq{\,\raise 0.3 ex\hbox{$ < $}\kern -0.75 em
 \lower 0.7 ex\hbox{$\sim$}\,}
 \def\gtsimeq{\,\raise 0.3 ex\hbox{$ > $}\kern -0.75 em
  \lower 0.7 ex\hbox{$\sim$}\,}

\def\avg #1{\langle #1\rangle}
\begin{frontmatter}

\title{Transition from Pareto to Boltzmann-Gibbs behavior in a deterministic economic model}
\author{\small\bf J. Gonz\'alez-Est\'evez$^{1,2}$, M. G. Cosenza$^2$,  O. Alvarez-Llamoza$^{2,3}$, and R. L\'opez-Ruiz$^4$}
\address{$^1$Laboratorio de F\'isica Aplicada y Computacional,  
Universidad Nacional Experimental del T\'achira, San Crist\'obal, Venezuela.\\
$^2$Centro de F\'isica Fundamental,  Universidad de Los Andes, M\'erida, Venezuela.\\
$^3$Departamento de F\'{\i}sica, FACYT, 
Universidad de Carabobo, Valencia, Venezuela. \\
$^4$DIIS and BIFI, Facultad de Ciencias, 
Universidad de Zaragoza, E-50009 Zaragoza, Spain.}

\date{\today}

\begin{abstract}
The one-dimensional deterministic economic model recently studied by Gonz\'alez-Est\'evez et al. 
[Physica A \textbf{387}, 4367 (2008)] is considered on a two-dimensional square lattice with periodic boundary conditions. In this model, the evolution of each agent is described by a map coupled with its nearest neighbors. The map has two factors: a linear term that accounts for the
agent's own tendency to grow and an exponential term that saturates this growth through 
the control effect of the environment. The regions in the parameter space where the system displays Pareto and Boltzmann-Gibbs statistics are calculated for the cases of von Neumann and of Moore's neighborhoods. It is found that, even when the parameters in the system are kept fixed, a transition from Pareto to Boltzmann-Gibbs behavior can occur when the number of neighbors of each agent increases.
\end{abstract}
\begin{keyword}
Multi-agent systems. Economic models. Pareto and Boltzmann-Gibbs distributions.
\PACS: 89.75.-k, 87.23.Ge, 05.90.+m  
\end{keyword}
\end{frontmatter}

\section{Introduction}

In the last few years, different probabilistic models \cite{yakovenko2000,chakraborti2000,chakraborti2004,angle2006,lopezruiz2007} have been proposed to explain the wealth distribution in western societies \cite{Dragulescu01,yakovenko2001,Chatterjee05,Chakrabarti06,yakovenko2007,Chatterjee07}, namely the Boltzmann-Gibbs distribution grouping about the $95\%$ of individuals, corresponding to those belonging to the low and middle economic classes, and the Pareto distribution consisting of the $5\%$ of individuals possessing the highest wealths. The majority of these models explain the wealth distribution as the consequence of random processes \cite{yakovenko2007}. However, some misconceptions lie along this line of research, since most people are aware that almost every economic trade is done under the rationale force of some interest or some final profit, and only occasionally this is conducted by chance. Thus, in order to shed light on the problem of how the wealth distribut!
 es in human society, it is important to dispose of other economic models incorporating different degrees of determinism in the interaction among the agents. 

One of these models was recently proposed \cite{sanchez2007} and studied in detail for the one-dimensional case in \cite{gonzalez2008}. This is a completely deterministic model that reproduces realistic wealth distributions, i.e., the Pareto and the Boltzmann-Gibbs (BG) distributions, for different values of parameters. Moreover, it is possible to produce a transition from Pareto to BG behavior by only modifying one parameter. It means that it is possible to bring the system from a particular wealth distribution to another one with a lower inequality by performing only a small change in the system configuration. This is an advantage respect to other random models where it is necessary to perform a structural reconfiguration of the system in order to get this type of transition \cite{chakraborti2000,chakraborti2004}. 

In this paper, we ask if some other strategies can be implemented in this model \cite{sanchez2007, gonzalez2008} in order to induce a transition in its asymptotic statistical behavior. We proceed to affirmatively answer this question by showing that different local groupings of the agents that occur, for instance, by changing the topology of the lattice or the number of neighbors of each agent, can lead to a transition from the Pareto to BG behavior
even when the parameters of the system are kept fixed. Hence, the first step we undertake in Section~2 is the characterization of the statistical behaviors that this model exhibits on its space of parameters when it is implemented on a two-dimensional lattice. Particularly, the cases of von Neumann and Moore neighborhoods are studied in detail. Then, in Section~3, the results obtained for these configurations are compared with those previously found for the one-dimensional case. The regions on the parameter space where a transition between different statistical behaviors can take place, are identified in this section. Section~4 contains our conclusions.
 
\section{The deterministic economic model on a two-dimensional lattice}

The system \cite{sanchez2007,gonzalez2008} consists of $N$ agents placed at the nodes of a network. Here, the network is a square lattice with periodic boundary conditions. Each agent, representing an individual, a company, a country or other economic entity, is identified by a pair of indexes $(i,j)$, with $i,j=1,\ldots,N$. The dynamics of each agent is described by a discrete-time map that expresses the competition between its own tendency to grow and an environmental influence that controls this growth. The dynamics of the system is described by the coupled map equations
\begin{equation}
\begin{array}{ll}
x_{t+1}^{i,j} = & r_{i,j}\: x_t^{i,j}\: \exp(-\mid x_t^{i,j}-a_{i,j}\Psi_t^{i,j}\mid), \\
\Psi_t^{i,j}=&\frac{1}{\eta(i,j)} \sum\limits_{i,j\in\nu(i,j)} x_{t}^{i,j},
\end{array}
\label{eq:system}
\end{equation}
where $x_t^{i,j}\geq 0$ gives the state of the agent $(i,j)$ at discrete time $t$, and it may denote the {\it wealth} of this agent; the factor  $r_{i,j}x_t^{i,j}$ expresses the {\it self-growth capacity} of agent $(i,j)$, characterized by a parameter $r_{i,j}$; $\Psi_t^{i,j}$ represents the local field acting at the site $(i,j)$ at time $t$; $\nu(i,j)$ is the set of agents in the network coupled to agent $(i,j)$ and $\eta(i,j)$ is the cardinality of this set; and $a_{i,j}$ measures the coupling of agent $(i,j)$ with its neighborhood; it can also be interpreted as the {\it local environmental pressure} exerted on agent $(i,j)$ \cite{ausloos2003}. The negative exponential function acts as a {\it control factor} that limits this growth with respect to the local field. With the dynamics given by Eqs.(\ref{eq:system}) the largest possibility of growth for agent $(i,j)$ is obtained when $x_t^{i,j}\simeq a_{i,j}\Psi_t^{i,j}$, i.e., when the agent has reached some kind of adaptatio!
 n to its local environment.

We consider two different nearest-neighbor interactions: the 4-cell von Neumann 
neighborhood, 
\begin{equation}
\Psi_t^{i,j} = \frac{1}{4} \left( x^{i-1,j}_t+x^{i+1,j}_t+x^{i,j-1}_t+x^{i,j+1}_t \right);
\end{equation}
and the 8-cell Moore neighborhood,
\begin{equation}
\begin{array}{ll}
\Psi_t^{i,j} = & \displaystyle{\frac{1}{8}} \displaystyle{(} x^{i-1,j}_t+x^{i+1,j}_t+x^{i,j-1}_t+x^{i,j+1}_t+ \\
              & x^{i-1,j-1}_t+x^{i-1,j+1}_t+x^{i+1,j-1}_t+x^{i+1,j+1}_t \displaystyle{)} \,.
\end{array}
\end{equation}

We focus on a homogeneous system where all agents possess the same growth capacity, $r_{i,j}=r$, and are subject to a uniform selection pressure from their environment,  $a_{i,j}=a$. Thus, the parameter $a$ can be interpreted as a homogeneous social constraint on the agents to reach a wealth level proportional to that of their environment. The value $a=1$ implies a tendency towards being totally balanced with the neighborhood. The cases $a<1$ corresponds to a lower level of expectation among agents that restricts the possibilities for improving their relative wealth. When $a>1$ the agents possess a greater stimulus for overcoming their local neighbors.

We study the collective behavior of the system described by Eqs.(\ref{eq:system}) in the space of parameters $(a,r)$. For all the simulations shown the system size is $316\times 316$ ($N\simeq 10^4$) and the values of the initial conditions are uniformly distributed at random in the interval $x^{i,j}_0 \in [1,100]$. Also, a transient of $10^4$ iterations is discarded before arriving to the asymptotic regime where all the calculations are carried out. When indicated, time averages are done over the next $100$ iterations after the transient, and this result is newly averaged  over $100$ different realizations of the initial conditions with the same parameter values.

\begin{figure}[t]
\centerline{\includegraphics[width=6.8cm]{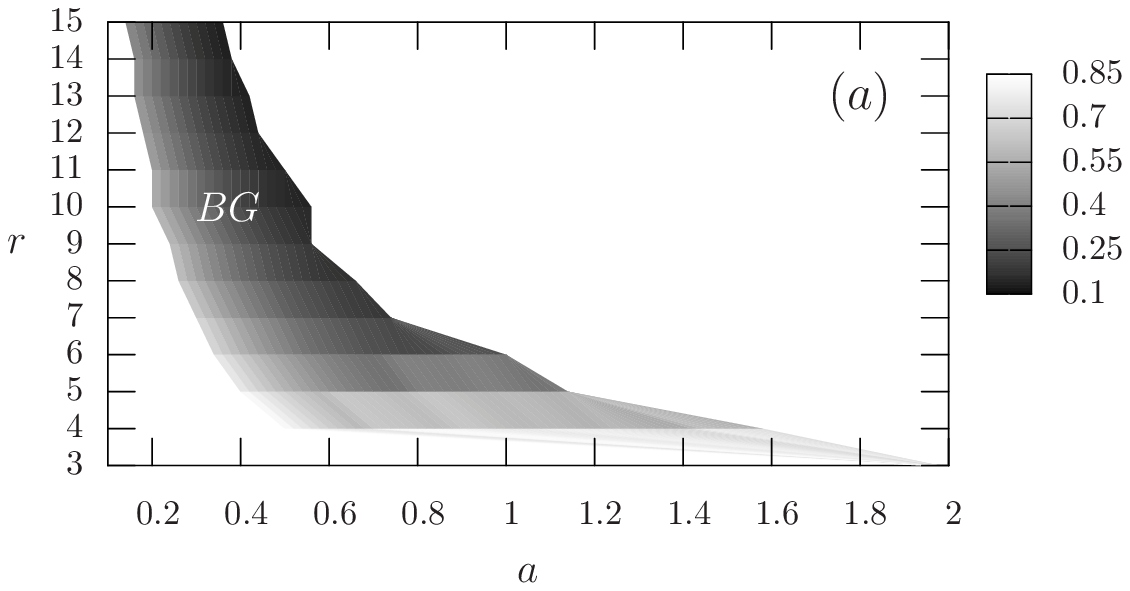}
\hspace{0mm}  
	    \includegraphics[width=6.8cm]{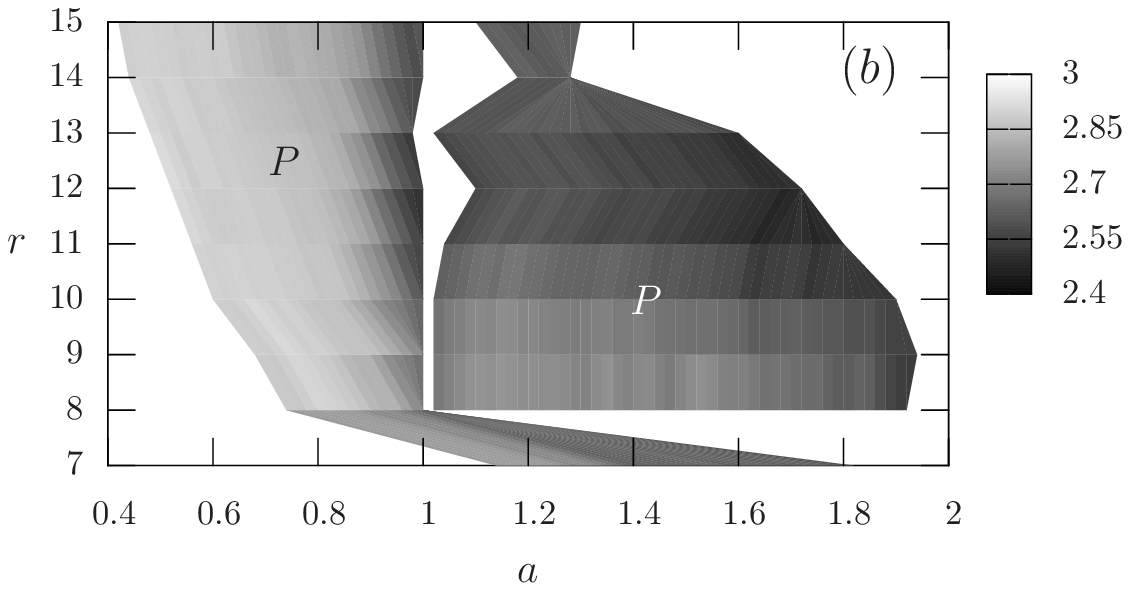}}
\vspace{10mm}
\centerline{\includegraphics[width=6.8cm]{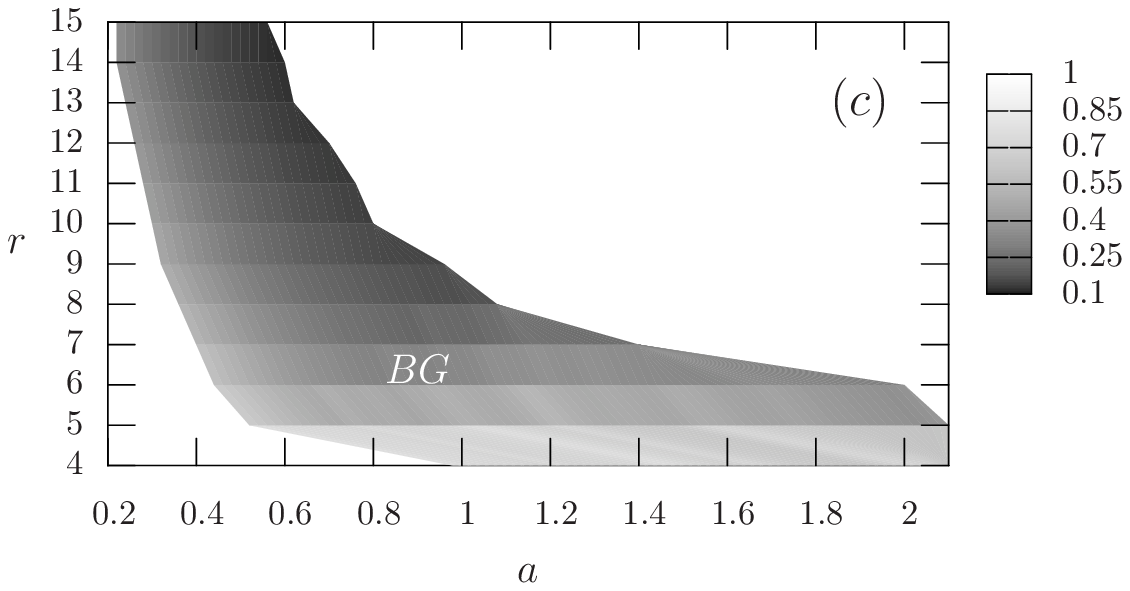}
\hspace{0mm}
            \includegraphics[width=6.8cm]{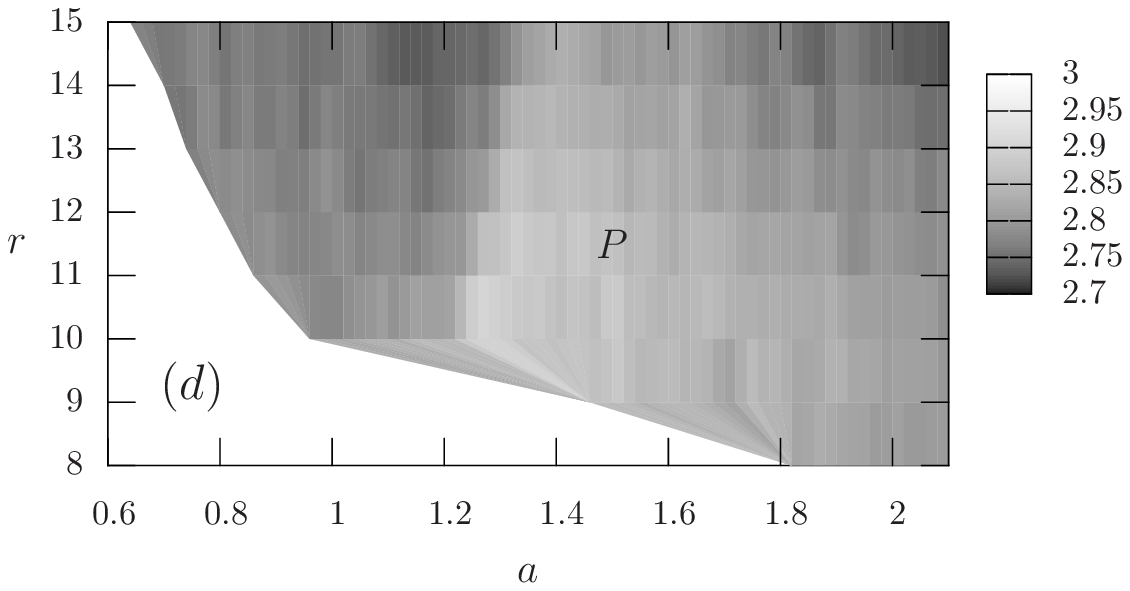}}
\vspace{0mm}
\caption{{\bf Left panels:} Regions where Boltzmann-Gibbs behavior $P(x)\sim e^{-\mu x}$ appears on the space of parameters $(a,r)$, indicated by the label \textit{BG}, for the {\bf (a)} 4-cell and {\bf (c)} 8-cell neighborhoods cases. The grey code on the right indicates the values of the scaling exponent $\mu$. {\bf Right panels:} Regions where Pareto behavior $P(x)\sim x^{-\alpha}$ occurs on the plane $(a,r)$, labeled by \textit{P}, for the {\bf (b)} 4-cell and {\bf (d)} 8-cell neighborhoods cases. The grey code on the right indicates the values of the exponent $\alpha$.}
\label{fig1}
\end{figure}

Figure~1 shows the regions where the probability distribution $P(x)$ displays BG behavior, $P(x)\sim e^{-\mu x}$, and Pareto behavior, $P(x)\sim x^{-\alpha}$, in the space of parameters $(a,r)$ for the system Eq.~(\ref{eq:system}). The parameters $a$ and $r$ are varied in intervals of size $\Delta a = 0.02$ and $\Delta r = 1$, respectively. For each pair $(a,r)$ and after undergoing the averaging process described above, semilog and log-log linear regressions are used to calculate the exponents $\mu$ and $\alpha$
for the corresponding distributions, and only those results yielding a correlation coefficient greater than $0.96$ are shown in Fig.~1.

In comparison with the one-dimensional case \cite{gonzalez2008}, the BG and Pareto regions in the space of parameters are larger for the square lattice. In both cases, the BG behavior occurs  for the lower values of the parameter $a$, but in two-dimensions there is also a narrow BG region for greater values of $a$. In general, when the value of $a$ increases, the population enters in a competitive regime that provokes the appearance of the Pareto behavior. The scaling exponents obtained for both regions are similar to the one-dimensional case. Those exponents that correspond to the Pareto behavior are in the range $\alpha\in [2.4,3.0]$, in agreement with those
found in actual economic data \cite{yakovenko2001,levy1997,souma2001}.

\begin{figure}[t]
\centerline{\includegraphics[width=6.8cm]{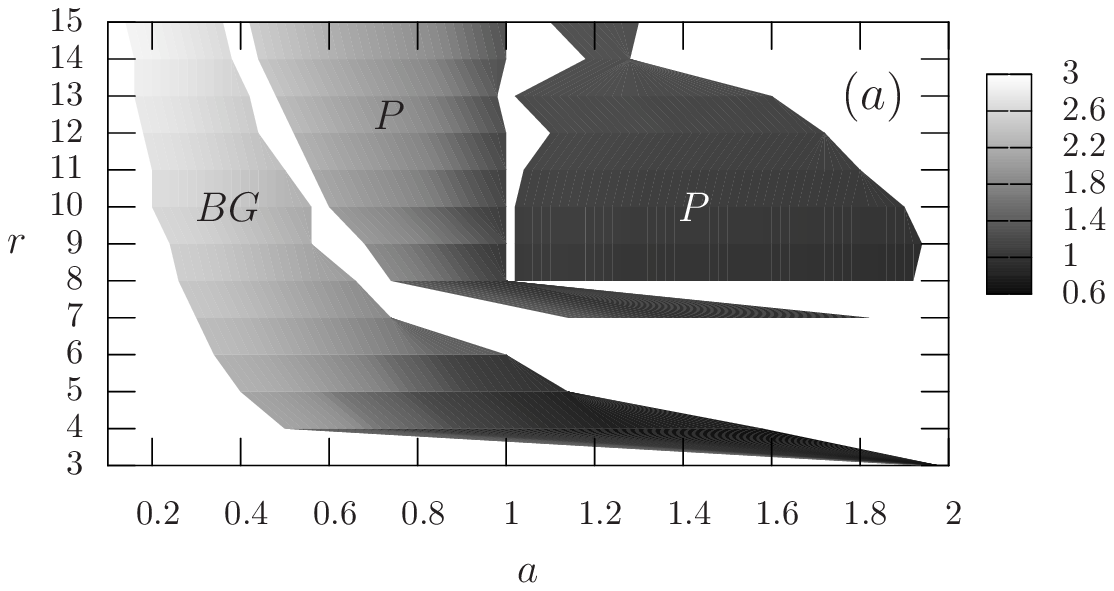}
\hspace{0mm}
            \includegraphics[width=6.8cm]{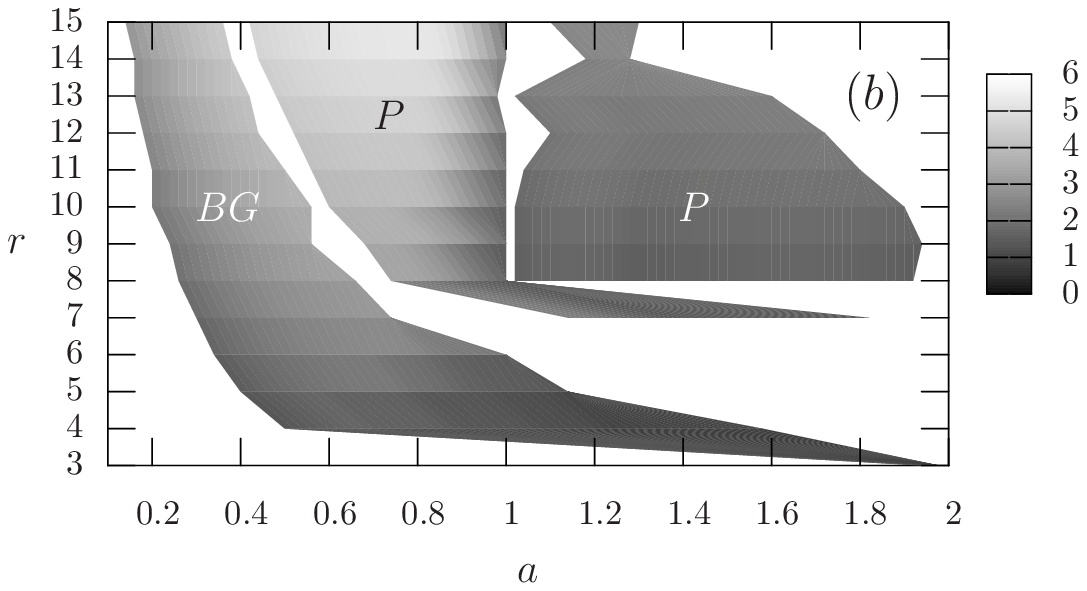}}
\vspace{10mm}
\centerline{\includegraphics[width=6.8cm]{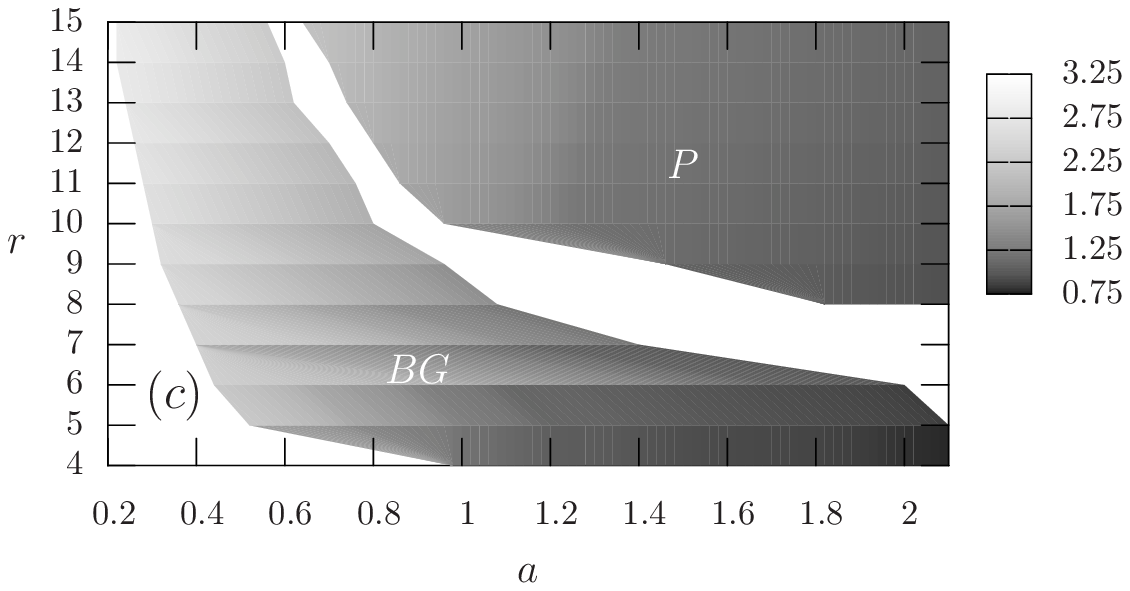}
\hspace{0mm}
            \includegraphics[width=6.8cm]{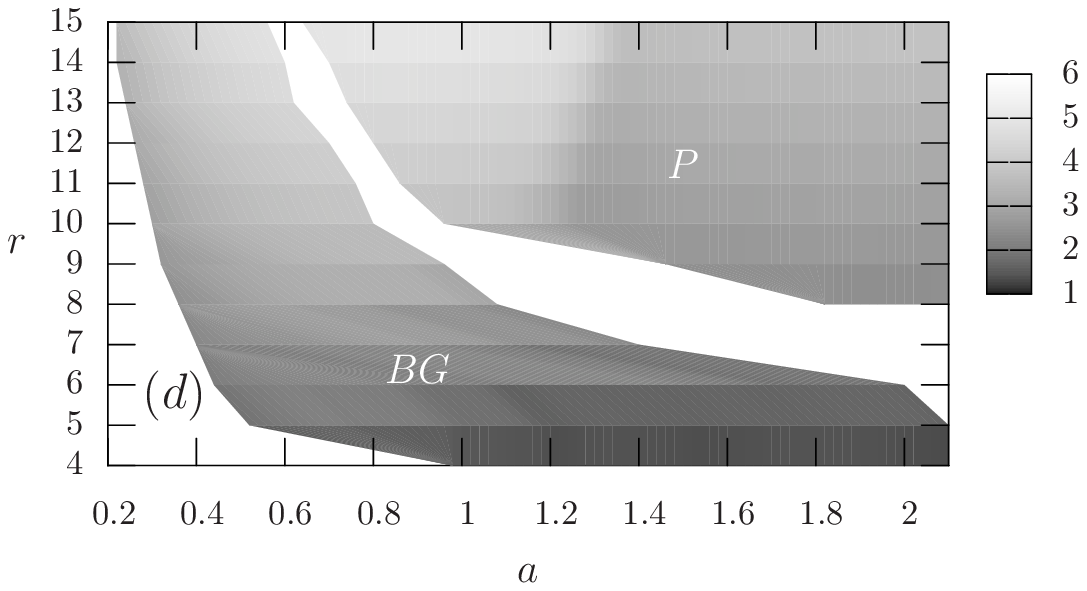}}
\caption{{\bf Left panels:} Mean field $H_t$ of the system at $t=10^4$ for the Boltzmann-Gibbs (BG) and Pareto (P) regions for the {\bf (a)} 4-cell and {\bf (c)} 8-cell neighborhoods cases. 
The grey code on the right indicates the values taken by $H_t$. {\bf Right panels:} The quantity $\langle\sigma\rangle$ on the space of parameters $(a,r)$, for the {\bf (b)} 4-cell and {\bf (d)} 8-cell neighborhoods cases, with the grey code indicated on the right.}
\label{fig2}
\end{figure}

The mean field of the system or average wealth per agent at a time $t$ is defined as 
\begin{equation}
H_{t}=\frac{1}{N}\sum_{i,j=1}^{N} x_t^{i,j}.
\end{equation} 
Similarly to the results reported in \cite{gonzalez2008} for the one-dimensional case, Figure~2(a-c) shows the asymptotic value of the global mean field for the regions where BG and Pareto behaviors are observed on the space of parameters $(a,r)$, for the 4-cell and 8-cell neighborhood cases. Note that, although the values of the initial states of the agents are randomly distributed on the interval $[1,100]$, the system evolves to an asymptotic state where $H_t$ takes values on the smaller interval $[0.6,3.25]$. Let us observe that the BG region presents a higher mean wealth than the Pareto one, telling us that in this model the BG behavior would be desirable for the general benefit of the ensemble. In the same line, the BG and Pareto regions with 8-cell neighborhood display a slightly bigger mean wealth than the 4-cell neighborhood and the 1D cases. It seems to indicate that the increasing number of the local neighbors of each agent can generate more richness in the system.!
  On the other hand, for some values of the parameters $a$ and $r$, the states of agents in the system at a given time exhibit a large dispersion. For those parameters, the values of the state of any agent present large fluctuations over long times. To characterize these fluctuations, we define the instantaneous standard deviation of the mean field as
\begin{equation}
\sigma_{t}=\left(\frac{1}{N}\sum_{i,j=1}^N\left[x_t^{i,j}-H_t \right]^{2} \right)^{1/2}.
\end{equation} 
After discarding $10^4$ transients, we calculate the mean value of $\sigma_t$  over $100$ iterations, and then average this result over $100$ realizations of initial conditions. The resulting average dispersion, denoted by $\langle  \sigma \rangle$, is shown in Fig.~2(b-d) on the plane $(a,r)$. Note that for some regions of parameters the quantity $\left\langle\sigma_t\right\rangle$ can be much greater than the value of the mean field. 

The large dispersions observed in Fig.~2 (b-d) reflect the inequality in the wealth distribution among agents in this system. To characterize the degree of inequality in the wealth distribution we use the Gini coefficient defined at a time $t$ as \cite{gini}
\begin{equation}
G_t=\frac{1}{2N^2 H_t} \sum_{i,j,k,l=1}^N |x_t^{i,j} - x_t^{k,l}|.
\end{equation}

\begin{figure}[h]
\centerline{\includegraphics[width=6.8cm]{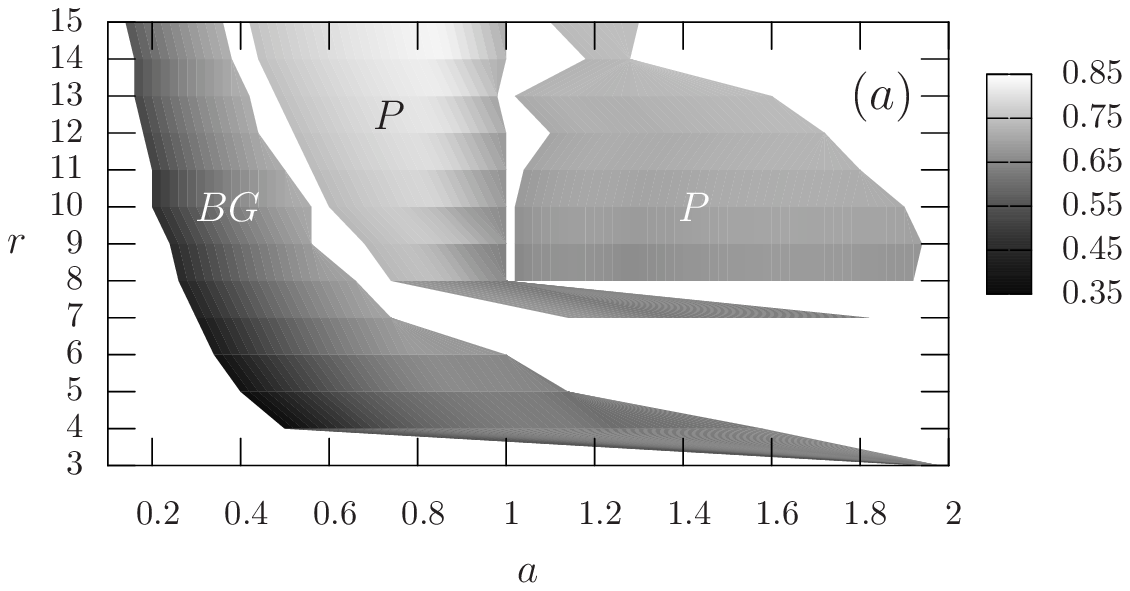}
\vspace{0mm}
            \includegraphics[width=6.8cm]{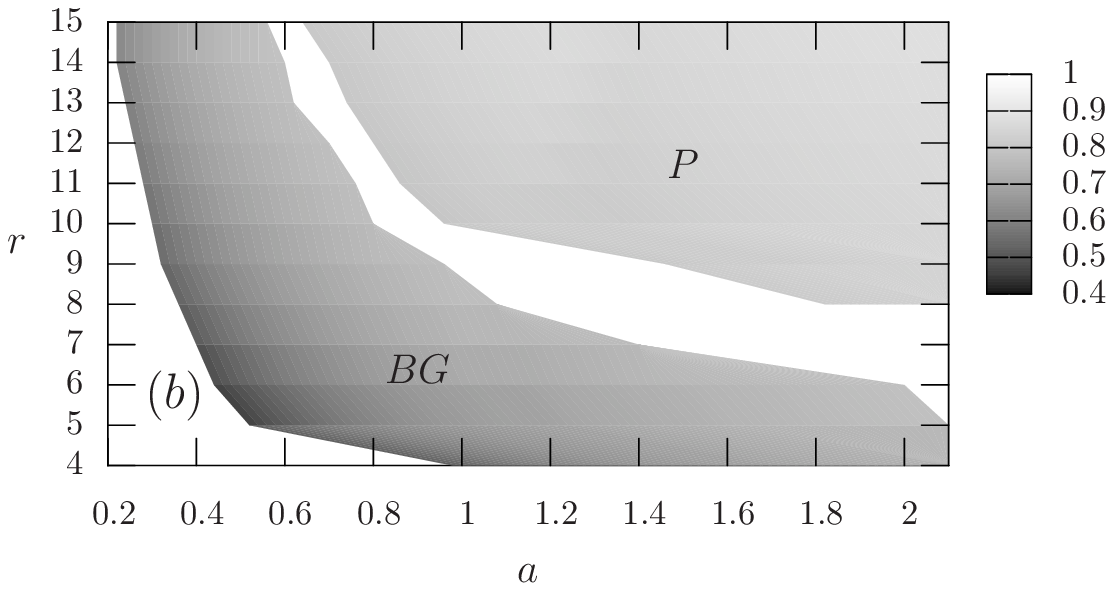}}
\caption{(a) Gini coefficient at $t=10^4$ as a function of the parameters $a$ and $r$ for the {\bf (a)} 4-cell and {\bf (b)} 8-cell neighborhoods cases. The labels \textit{BG} and \textit{P} indicate the regions with Boltzmann-Gibbs and Pareto behaviors. The grey code on the right indicates the values taken by the Gini coefficient.}
\label{fig3}
\end{figure}

A perfectly equitable  distribution of wealth at time $t$, where $x_t^{i,j}=x_t^{k,l} \; \forall \; i,j,k,l$, yields a value $G_t=0$. The opposite situation, where one agent has the total wealth $\sum\limits_{i,j=1}^N x_t^{i,j}$, has a value of $G_t=1$. Figure~3 shows the asymptotic value of the Gini coefficient on the plane of parameters $(a,r)$. Note that the Gini coefficient reaches larger values, i.e.  $G_t \in [0.65,0.85]$ and $G_t \in [0.7,1]$ for the 4-cell and 8-cell neighborhoods, respectively, in the regions associated to Pareto regimes, while it takes lower values, i.e. $G_t \in[0.35,0.65]$, in the region corresponding to BG behavior. This results agree with our qualitative understanding that equity is more favored in the presence of a larger middle economic class in a society, as expressed by a BG distribution. 

\section{Transition from Pareto to Boltzmann-Gibbs behavior}

As it was put in evidence in the one-dimensional realization of this economic system studied in \cite{gonzalez2008}, it is possible to produce a transition from a Pareto to a BG behavior (or viceversa) maintaining its structural configuration and performing only a variation in the parameters $(a,r)$ of the system. In the two-dimensional case, as it can be seen in Fig.~1, it is also possible to make such transition, for example, by changing $r$  with a fixed value of $a$. We call the transition from Pareto to BG behavior (or viceversa) induced by varying the parameters the {\it Transition Strategy I}.
  
Figure~4 reveals another mechanism capable of provoking a change in the statistical properties of this economic model. We call it the {\it Transition Strategy II}. It consists in performing a reconfiguration of the local connectivity of the agents. Figure~4(a) shows the region of the parameter space $(a,r)$ where a change of the topology of system from a one-dimensional array to a two-dimensional square lattice with 4-cell neighborhood takes the system from a Pareto to a BG asymptotic distribution. The same transition can be reached in the region plotted in Fig.~4(b)
by transforming the one-dimensional array to a two-dimensional lattice with 8-cell neighborhood. Figure~4(c) displays the region of parameter space where a modification of the number of neighbors, from a 4-cell von Neumann neighborhood to a 8-cell Moore neighborhood, generates a Pareto-BG transition in the two-dimensional array. 

As shown in Fig.~4(d), the transitions in statistical behaviors achieved through either type of strategy produce a decrease of the Gini coefficient and, therefore, a more equitable wealth distribution in the system. 

\begin{figure}[h]
\centerline{\includegraphics[width=6.8cm]{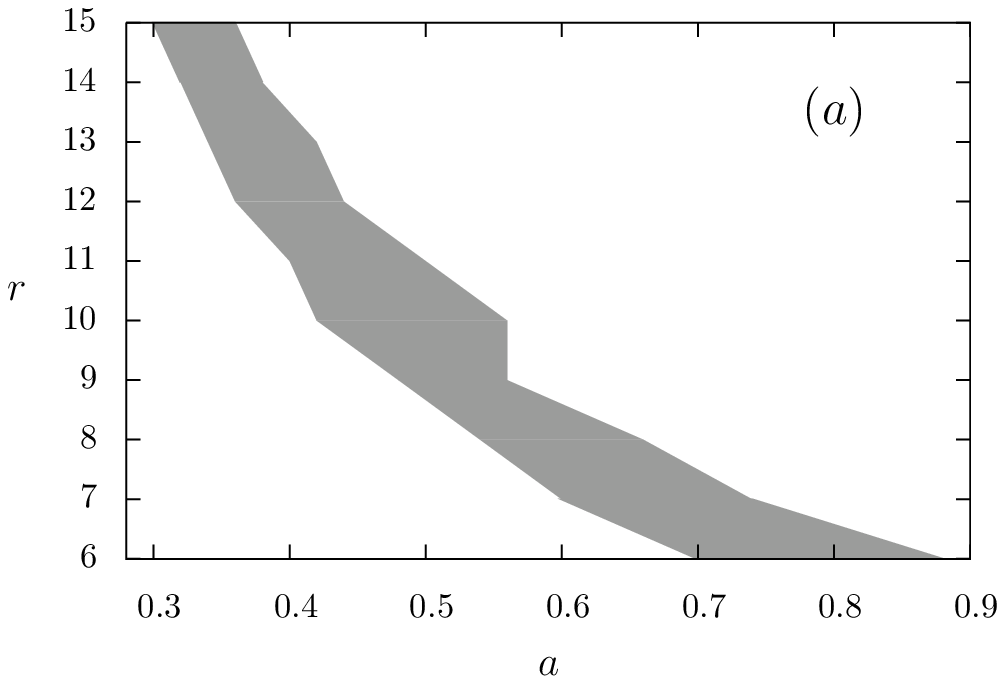}
\hspace{0mm}
            \includegraphics[width=6.8cm]{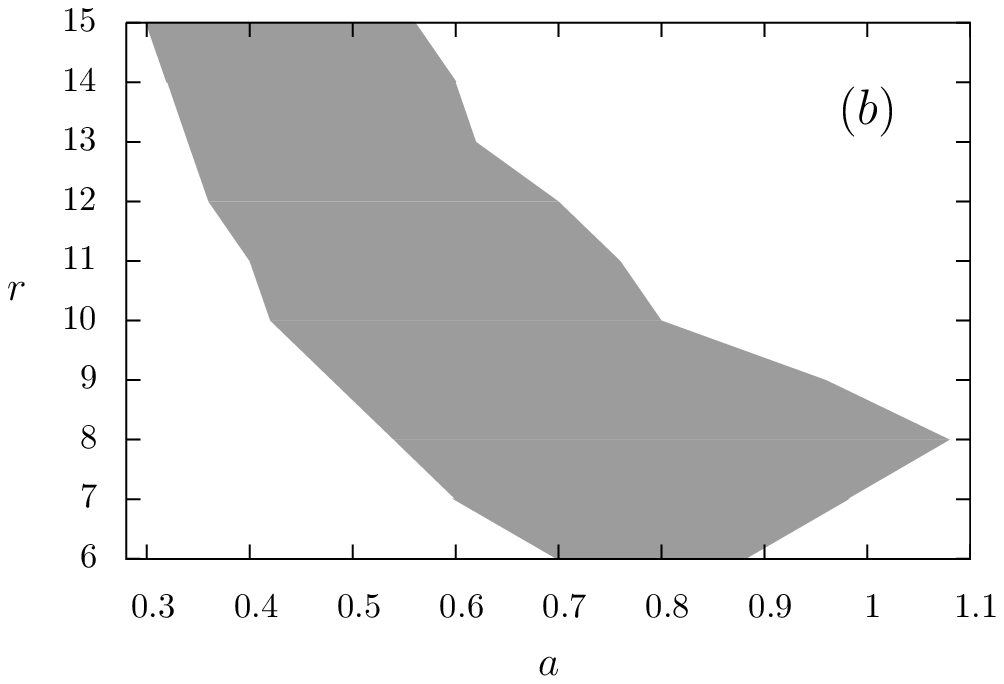}}
\vspace{10mm}
\centerline{\includegraphics[width=6.8cm]{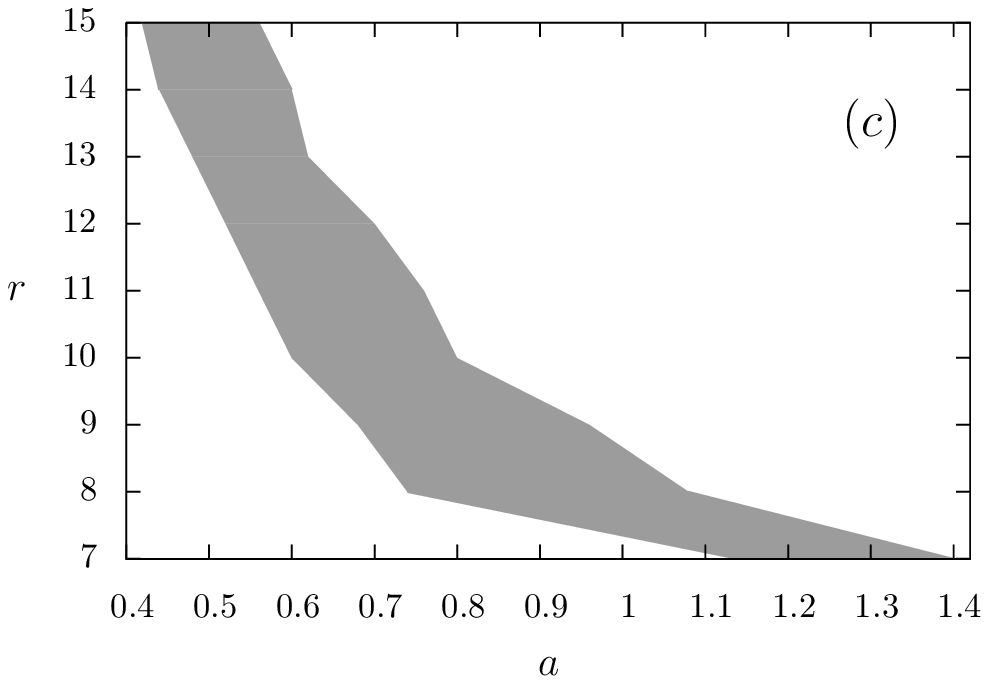}
\hspace{0mm}
            \includegraphics[width=6.8cm,height=4.6cm]{fig4d.eps}}
\vspace{0mm}
\caption{Regions in the parameter space $(a,r)$ where the deterministic model displays a transition from Pareto to Boltzmann-Gibbs statistical behavior through a reconfiguration of the local neighborhoods. (a) The system changes its topology from a one-dimensional array to a two-dimensional square lattice with a 4-cell neighborhood. (b)  The topology changes from a one-dimensional array to a two-dimensional square latticeto a 8-cell neighborhood. 
(c) Changing from a 4-cell neighborhood to a 8-cell neighborhood in a two-dimensional square lattice. (d) Graphical evidence that the Gini coefficient decreases in the transitions described in (a), (b), and (c): one-dimensional array (circle points, upper line); two-dimensional lattice with 4-cell neighborhood (square points, middle line);  8-cell neighborhood (triangle down points, lower line). Fixed $r=10$.}
\label{fig4}
\end{figure}

\section{Conclusions}

Nowadays, both the exponential and power law statistical behaviors in the economies of western countries are well documented. Different models for the interaction of the economic agents have been reported in the literature. Those models with a probabilistic inspiration must apply a strong reconfiguration in the structural properties of the system in order to obtain a transition from a BG to a Pareto distribution in its asymptotic state.

In this paper we have characterized in detail a deterministic economic model implemented on a two-dimensional square lattice with periodic boundary conditions. The system depends only on two parameters; the parameter $a$ brings information about the local environmental pressure while $r$ expresses the self-growth capacity of each agent. By varying the values of these parameters, the model can exhibit asymptotic BG or Pareto distributions of wealth. The striking fact is that only 
a change in either one of these parameters is enough to make the system undergo from one type of statistical behavior to a different one. We have called this process Transition Strategy I.

Another situation that can produce a transition from Pareto to BG behavior in this system has been found in this work. We have called it Transition Strategy II. In this scenario, a variation of the values of the parameters is not necessary; the transition in the type of asymptotic statistical behavior is a consequence of the rearrangement of the neighborhoods of each agent. In particular, it is observed that an increase in the number of neighbors of each agent drives the system toward asymptotic states where the wealth is distributed more equitatively, that is, with a lower value of the Gini coefficient. The specific regions of the parameter space where this transition can take place have been delimited.

To our knowledge this is the first time in the field of economic systems where this kind of transitions is reported. We hope that this work can help to enlighten the mechanisms that operate in the complex world of economical transactions among individuals, companies, countries or other economic entities.

\section*{Acknowledgments}

This work was supported in part by Decanato de Investigaci\'on of the Universidad Nacional Experimental del T\'achira (UNET), under grants 04-001-2006 and 04-002-2006. R.L-R. acknowledges financial support from Postgrado de F\'{\i}sica Fundamental, Universidad de Los Andes (ULA), Venezuela, and by grant DGICYT-FIS2006-12781-C02-01, Spain. He wants also to thank ULA and UNET for their kind hospitality during his stay there in July 2008. M. G. C. and O.A-L. acknowledge support from  Consejo de Desarrollo, Cient\'ifico, Tecnol\'ogico y Human\'istico, ULA, under grant C-1579-08-05-B.

\end{document}